\documentclass[english,aps,preprint]{revtex4}
\usepackage[T1]{fontenc}
\usepackage[latin9]{inputenc}
\usepackage{array}
\usepackage{longtable}
\usepackage{graphicx}
\usepackage{amssymb}

\providecommand{\tabularnewline}{\\}

\usepackage{babel}

\begin{document}

\title{Anomalous Single Top Production at the LHeC Based $\gamma$p Collider}

\author{\.{I}.T. Çakir}

\email{tcakir@mail.cern.ch}

\affiliation{Department of Physics, CERN, Geneva, Switzerland}

\author{O. Çak\i{}r }

\email{ocakir@science.ankara.edu.tr}

\affiliation{Ankara University, Faculty of Sciences, Department of Physics, Ankara,
Turkey}

\author{S. Sultansoy}

\email{ssultansoy@etu.edu.tr}

\affiliation{Physics Division, TOBB University of Economics and Technology, Ankara,
Turkey}

\affiliation{Institute of Physics, Academy of Sciences, Baku, Azerbaijan}
\begin{abstract}
The top quark could provide very important information for the Standard
Model extentions due to its large mass close to the electroweak symmetry
breaking scale. In this work, anomalous single top production is studied
by using $\gamma p\rightarrow W^{+}b$ process at the LHeC based $\gamma p$
collider. The sensitivity to anomalous coupling $\kappa/\Lambda$
could be reached down to 0.01 TeV$^{-1}$.
\end{abstract}
\maketitle
The top quark is considered to be the most sensitive to the new physics
beyond the Standard Model (BSM) since it is the heaviest available
particle of the Standard Model (SM). If the BSM is associated with
the mass generation, the top quark interactions will be sensitive
to the mechanism of dynamical symmetry breaking. The precise measurement
of the couplings between SM bosons and fermions provides powerful
tool for the search of the BSM physics. As mentioned in \cite{1},
the effects of new physics on the top quark couplings are expected
to be larger than that on any other fermions, and deviations with
respect to the SM predictions might be detectable. 

A possible anomalous $tqV$ ($V=g,\gamma,Z$ and $q=u,c$) couplings
can be generated through a dynamical mass generation \cite{2}. They
have a similar chiral structure as the mass terms, and the presence
of these couplings would be interpreted as signals of new interactions.
This motivates the study of top quarks' flavour changing neutral current
(FCNC) couplings at present and future colliders. 

Current experimental constraints at 95\% C.L. on the anomalous top
quark couplings are \cite{3}: $BR(t\rightarrow\gamma u)<0.0132$
and $BR(t\rightarrow\gamma u)<0.0059$ from HERA; $BR(t\rightarrow\gamma q)<0.041$
from LEP and $BR(t\rightarrow\gamma q)<0.032$ from CDF. The HERA
has much higher sensitivity to $u\gamma t$ than $c\gamma t$ due
to more favorable parton density: the best limit is obtained from
the ZEUS experiment. 

The top quarks will be produced in large numbers at the Large Hadron
Collider (LHC), therefore the couplings of the top quark can be probed
with a great precision. For a luminosity of 1 fb$^{-1}$ the expected
ATLAS sensitivity to the top quark FCNC decay is $BR(t\to q\gamma)\sim10^{-3}$
at 95\% C.L. \cite{4}. For $L_{int}=$100 fb$^{-1}$ the ATLAS sensitivity
to $t\gamma q$ anomalous interactions has been estimated as $BR(t\to q\gamma)\sim10^{-4}$
at $5\sigma$ level \cite{5}. 

The production of top quarks by FCNC interactions at hadron colliders
has been studied in \cite{6}, $e^{+}e^{-}$colliders in \cite{2,7}
and lepton-hadron collider in \cite{2,8}. LHC will give an opportunity
to probe $BR(t\rightarrow ug)$ down to $5\times10^{-3}$ \cite{9};
ILC/CLIC has the potential to probe $BR(t\rightarrow q\gamma)$ down
to $10^{-5}$ \cite{10}.

It is known that linac-ring type colliders present the sole realistic
way to TeV scale in lepton-hadron collisions \cite{11}. An essential
advantage of linac-ring type ep-colliders is the opportunity to construct
$\gamma p$ colliders on their basis \cite{12}. Construction of linear
$e^{+}e^{-}$collider or special linac tangential to LHC ring will
give opportunity to utilize highest energy proton and nuclei beams
for lepton-hadron collisions. Recently this opportunity is widely
discussed in the framework of the LHeC project \cite{13}. Two stages
of the LHeC are considered: QCD Explorer ($E_{e}=50-100$ GeV) and
Energy Frontier ($E_{e}>250$ GeV). First stage is mandatory for two
reasons: to provide precision PDF's for adequate interpretation of
LHC data and to enlighten QCD basics.

In this paper, we investigate the potential of LHeC based $\gamma p$
collider to search for anomalous top quark interactions. 

The effective Lagrangian involving anomalous $t\gamma q$ $(q=u,c)$
interactions is given by \cite{9}

\begin{equation}
L=-g_{e}\sum_{q=u,c}{\displaystyle Q_{q}\frac{\kappa_{q}}{\Lambda}\bar{t}\sigma^{\mu\nu}(f_{q}+h_{q}\gamma_{5})qA_{\mu\nu}+h.c.}\label{eq:1}\end{equation}
 where $A_{\mu\nu}$ is the usual photon field tensor, $\sigma_{\mu\nu}=\frac{i}{2}(\gamma_{\mu}\gamma_{\nu}-\gamma_{\nu}\gamma_{\mu})$,
$Q_{q}$ is the quark charge, in general $f_{q}$ and $h_{q}$ are
complex numbers, $g_{e}$ is electromagnetic coupling constant, $\kappa_{q}$
is real and positive anomalous FCNC coupling and $\Lambda$ is the
new physics scale. The neutral current magnitudes in the Lagrangian
satisfy $|(f_{q})^{2}+(h_{q})^{2}|=1$ for each term. Using Eq. \ref{eq:1},
the anomalous decay width can be calculated as 

\begin{equation}
\Gamma(t\rightarrow q\gamma)=(\frac{\kappa_{q}}{\Lambda})^{2}\frac{2}{9}\alpha_{em}m_{t}^{3}\label{eq:2}\end{equation}

\begin{figure}
\includegraphics{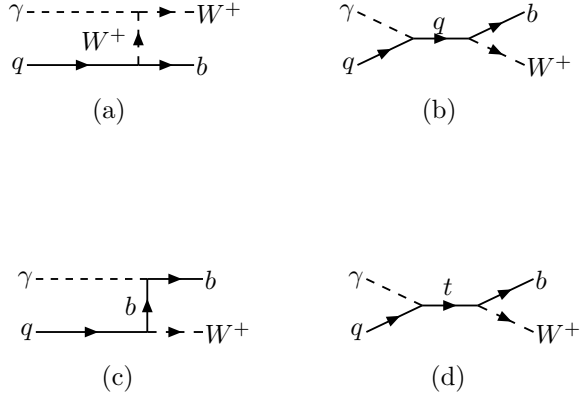}\caption{Feynman diagrams for $\gamma q\rightarrow W^{+}b$, where $q=u,c$.\label{fig:1}}

\end{figure}

Taking $m_{t}=173$ GeV and $\alpha_{em}=0.0079$, we find the anomalous
decay width $\approx9$ MeV for $\kappa_{q}/\Lambda=1$ TeV$^{-1}$,
while the SM decay width is about 1.5 GeV. For numerical calculations
we implemented anomalous interaction vertices from Lagrangian (\ref{eq:1})
into the CalcHEP package \cite{14} and use PDF library CTEQ6M \cite{15}.
The Feynman diagrams for the subproces $\gamma q\rightarrow W^{+}b$,
where $q=u,c$ are presented in Fig. \ref{fig:1}. First three diagrams
correspond to irreducible background and the last one to signal. 

The main background comes from associated production of $W$ boson
and the light jets. Hereafter, for b-tagging efficiency we used the
60\% and the mistagging factors for light ($u,d,s$) and $c$ quarks
are taken as 0.01 and 0.1, respectively. 

\begin{figure}
\includegraphics[scale=0.8]{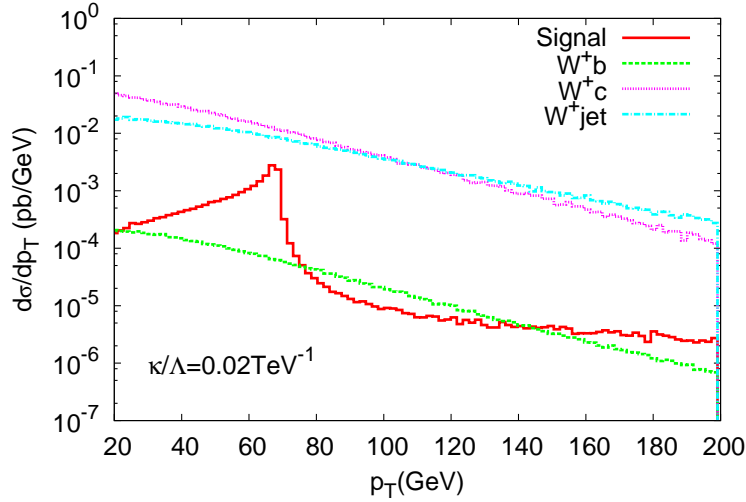}\caption{The transverse momentum distribution of the final state jet for the
signal and background processes. The differential cross section includes
the b-tagging efficiency and the rejection factors for the light jets.
Here the center of mass energy $\sqrt{s_{ep}}=1.4$ TeV and $\kappa/\Lambda=$0.02
TeV $^{-1}$. \label{fig:2}}

\end{figure}

\begin{figure}
\includegraphics[scale=0.8]{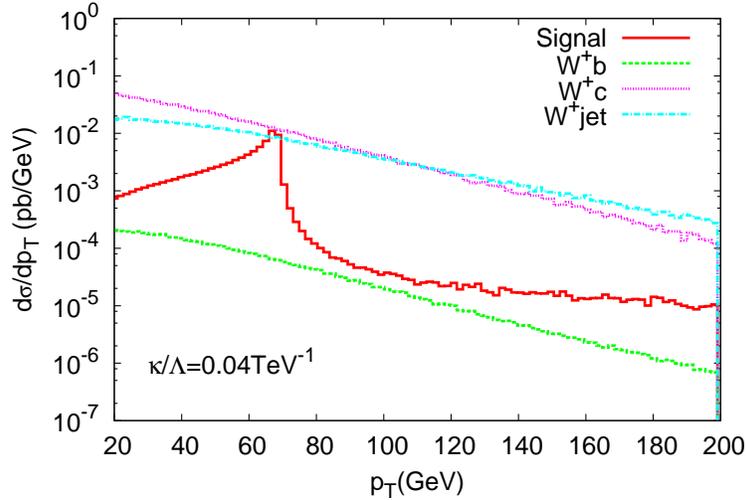}\caption{The same as Figure 2 but for $\kappa/\Lambda=0.04$ TeV$^{-1}$.\label{fig:3} }

\end{figure}

The differential cross sections of the final state jets are given
in Fig. \ref{fig:2} ($\kappa/\Lambda=0.02$ TeV$^{-1}$) and Fig.
\ref{fig:3} ($\kappa/\Lambda=0.04$ TeV$^{-1}$) for $E_{e}=70$
GeV and $E_{p}=7000$ GeV. Here, we assume $\kappa_{u}=\kappa_{c}=\kappa$.
The transverse momentum distribution of the signal has a peak around
70 GeV. 

\begin{figure}
\includegraphics[scale=0.8]{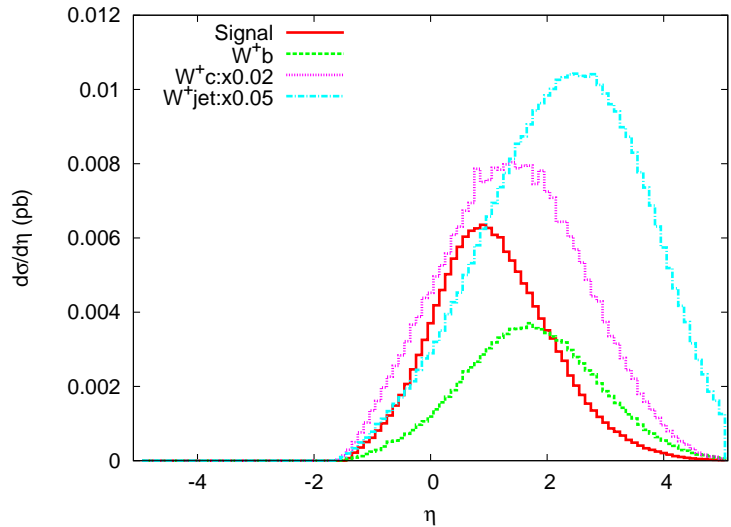}\caption{Pseudo-rapidity distribution of the jets in the signal ($\kappa/\Lambda=0.01$
TeV$^{-1}$) and background processes, where we applied a cut $p_{T}>20$
GeV. Here, $E_{e}=70$ GeV and $E_{p}=7000$ GeV. \label{fig:4}}

\end{figure}

The pseudo-rapidity distribution of the jets in the signal ($\kappa/\Lambda=0.01$
TeV$^{-1}$) and background processes are presented in Fig. \ref{fig:4},
where we applied a cut $p_{T}>20$ GeV. The maximum of the signal
is around $\eta=1$, while the main background shifted to $\eta\thicksim2$.
Nevertheless, one can see from Fig. \ref{fig:4} that $\eta$ cut
does not provide essential gain. 

The cross sections for signal and background processes with different
$p_{T}$ cuts are presented in Table \ref{tab:1}. It is seen that
$p_{T}$ cut slightly reduce the signal $(\sim30\%$ for $p_{T}>50$
GeV), whereas the background is essentially reduced (factor 4-6).
In order to improve the signal to background ratio further one can
use invariant mass ($W+jet$) cut around top mass. In Table \ref{tab:2},
the cross sections for signal and background processes are given using
both $p_{T}$ and invariant mass cuts $(M_{Wb}=150-200$ GeV).

\begin{table}
\caption{The cross sections (in pb) according to the $p_{T}$ cut for the signal
and background at $\gamma p$ collider based on the LHeC with $E_{e}=70$
GeV and $E_{p}=$7000 GeV. \label{tab:1}}

\begin{tabular}{|c|c|c|c|c|}
\hline 
$\kappa/\Lambda=0.01$ TeV$^{-1}$ & No cut & $p_{T}>20$ GeV & $p_{T}>40$ GeV & $p_{T}>50$ GeV\tabularnewline
\hline
\hline 
Signal & $9.54\times10^{-3}$ & $9.16\times10^{-3}$ & $7.84\times10^{-3}$ & $6.66\times10^{-3}$\tabularnewline
\hline 
Background: $W^{+}b$ & $9.60\times10^{-3}$ & $6.18\times10^{-3}$  & $3.48\times10^{-3}$ & $2.55\times10^{-3}$\tabularnewline
\hline 
Background: $W^{+}c$ & $3.11\times10^{0}$ & $1.27\times10^{0}$ & $6.85\times10^{-1}$ & $4.90\times10^{-1}$\tabularnewline
\hline 
Background: $W^{+}jet$ & $1.79\times10^{0}$ & $7.24\times10^{-1}$ & $4.79\times10^{-1}$ & $3.77\times10^{-1}$\tabularnewline
\hline
\end{tabular}
\end{table}

\begin{table}
\caption{The cross sections (in pb) according to the $p_{T}$ cut and invariant
mass interval $(M_{Wb}=150-200$ GeV) for the signal and background
at $\gamma p$ collider based on the LHeC with $E_{e}=70$ GeV and
$E_{p}=$7000 GeV. \label{tab:2}}

\begin{tabular}{|c|c|c|c|}
\hline 
$\kappa/\Lambda=0.01$ TeV$^{-1}$ & $p_{T}>20$ GeV & $p_{T}>40$ GeV & $p_{T}>50$ GeV\tabularnewline
\hline
\hline 
Signal & $8.86\times10^{-3}$ & $7.54\times10^{-3}$ & $6.39\times10^{-3}$\tabularnewline
\hline 
Background: $W^{+}b$ & $1.73\times10^{-3}$ & $1.12\times10^{-3}$ & $7.69\times10^{-4}$\tabularnewline
\hline 
Background: $W^{+}c$ & $3.48\times10^{-1}$ & $2.30\times10^{-1}$ & $1.63\times10^{-1}$\tabularnewline
\hline 
Background: $W^{+}jet$ & $1.39\times10^{-1}$ & $9.11\times10^{-2}$ & $6.38\times10^{-2}$\tabularnewline
\hline
\end{tabular}
\end{table}

In order to calculate the statistical significance (\emph{SS}) we
use following formula \cite{16} :

\begin{equation}
SS=\sqrt{2\left[(S+B)\ln(1+\frac{S}{B})-S\right]}\label{eq:3}\end{equation}
where $S$ and $B$ are the numbers of signal and background events,
respectively. Results are presented in Table \ref{tab:3} for different
$\kappa/\Lambda$ and luminosity values. It is seen that even with
$2$ fb$^{-1}$ the LHeC based $\gamma p$ collider will provide $5\sigma$
discovery for $\kappa/\Lambda=0.02$ TeV$^{-1}$.

\begin{table}
\caption{The signal significance ($SS$) for different values of $\kappa/\Lambda$
and integral luminosity for $E_{e}=70$ GeV and $E_{p}=$7000 GeV
(the numbers in parenthesis correspond to $E_{e}=140$ GeV). \label{tab:3} }

\begin{longtable}{|c|l|c|}
\hline 
$SS$ & $L=2$ fb$^{-1}$ & $L=10$ fb$^{-1}$\tabularnewline
\hline 
\multicolumn{1}{|c|}{$\kappa/\Lambda=0.01$ TeV$^{-1}$} & 2.58 (2.88) & 5.79 (6.47)\tabularnewline
\hline 
$\kappa/\Lambda=0.02$ TeV$^{-1}$ & 5.26 (5.92) & 11.78 (13.25)\tabularnewline
\hline
\end{longtable}
\end{table}

\begin{figure}
\includegraphics[scale=0.6]{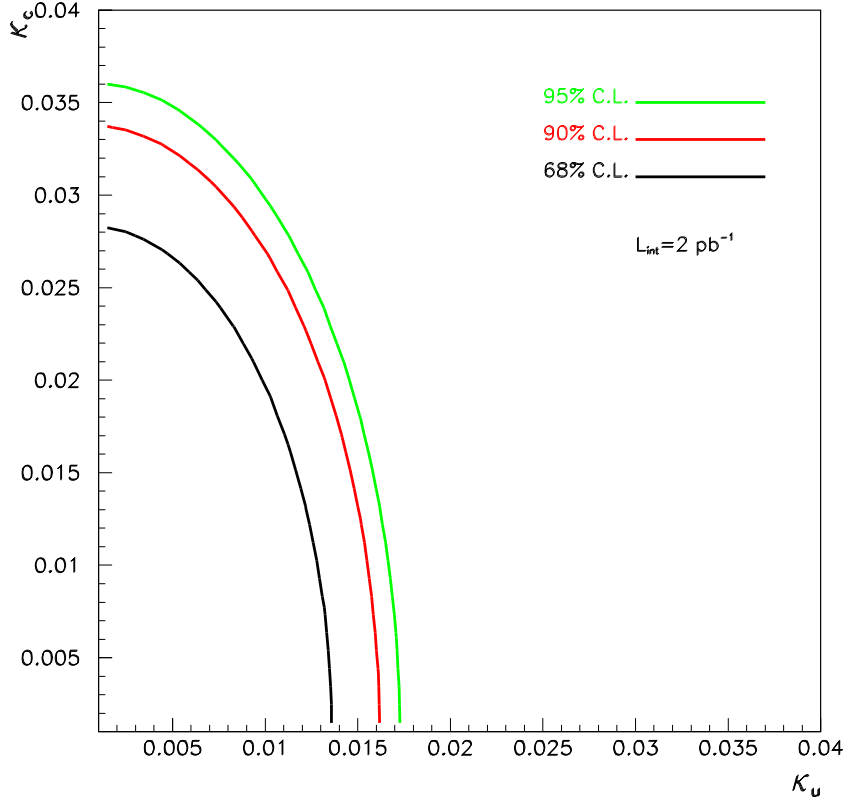}\caption{Contour plot for the anomalous couplings reachable at the LHeC based
$\gamma p$ collider with the ep center of mass energy $\sqrt{s_{ep}}=1.4$
TeV and integrated luminosity $L_{int}=2$ fb$^{-1}$. \label{fig:5} }

\end{figure}

\begin{figure}
\includegraphics[scale=0.6]{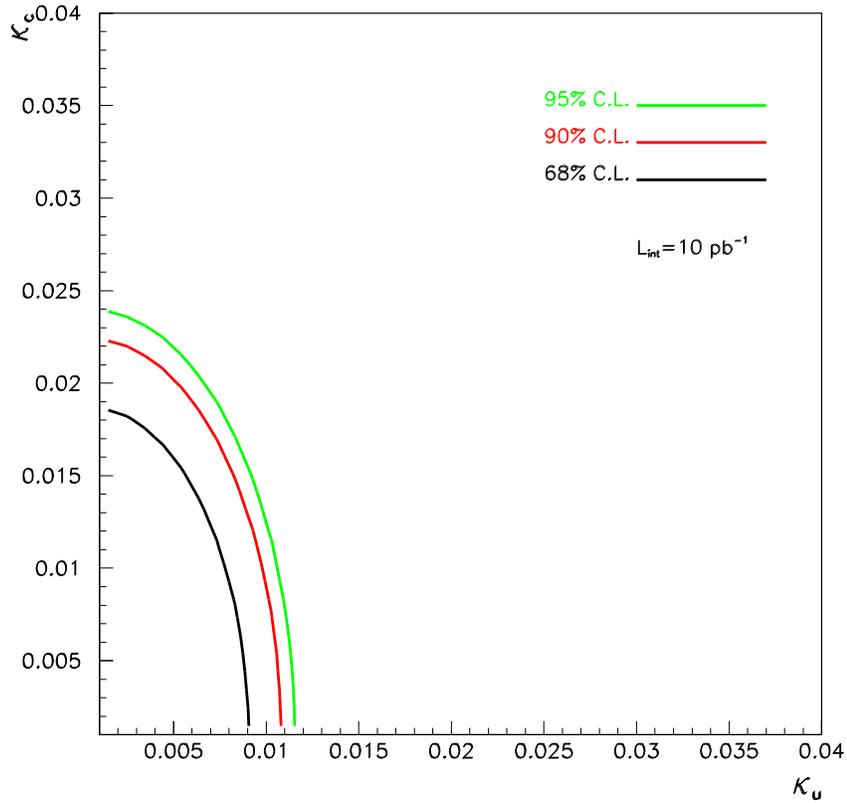}\caption{The same as Figure 5 but for $L_{int}=10$ fb$^{-1}$.\label{fig:6}}

\end{figure}

\begin{figure}
\includegraphics[scale=0.6]{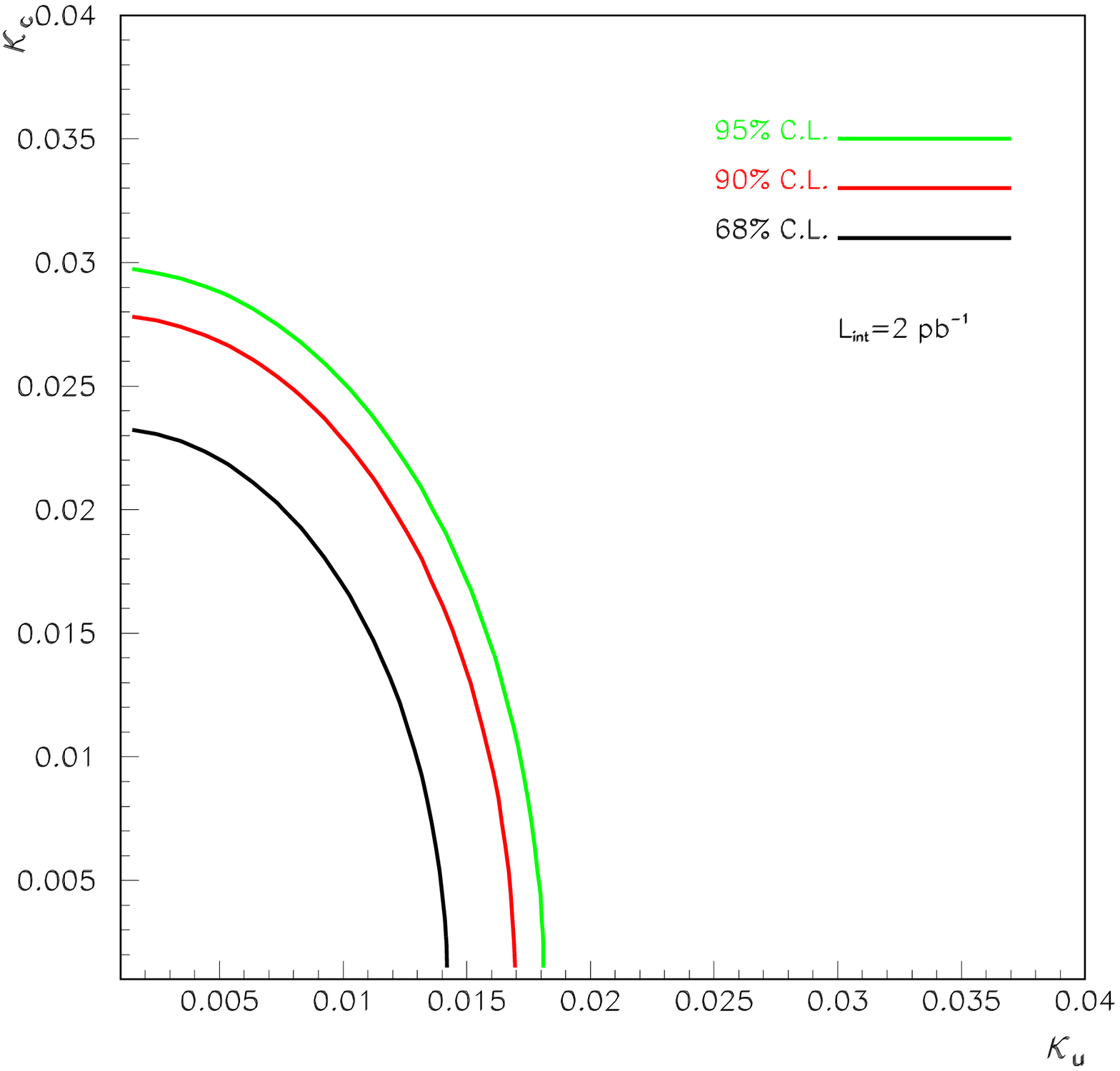}\caption{Contour plot for the anomalous couplings reachable at the LHeC based
$\gamma p$ collider with the ep center of mass energy$\sqrt{s_{ep}}=1.9$
TeV and integrated luminosity $L_{int}=2$ fb$^{-1}$ \label{fig:7}. }

\end{figure}

\begin{figure}
\includegraphics[scale=0.6]{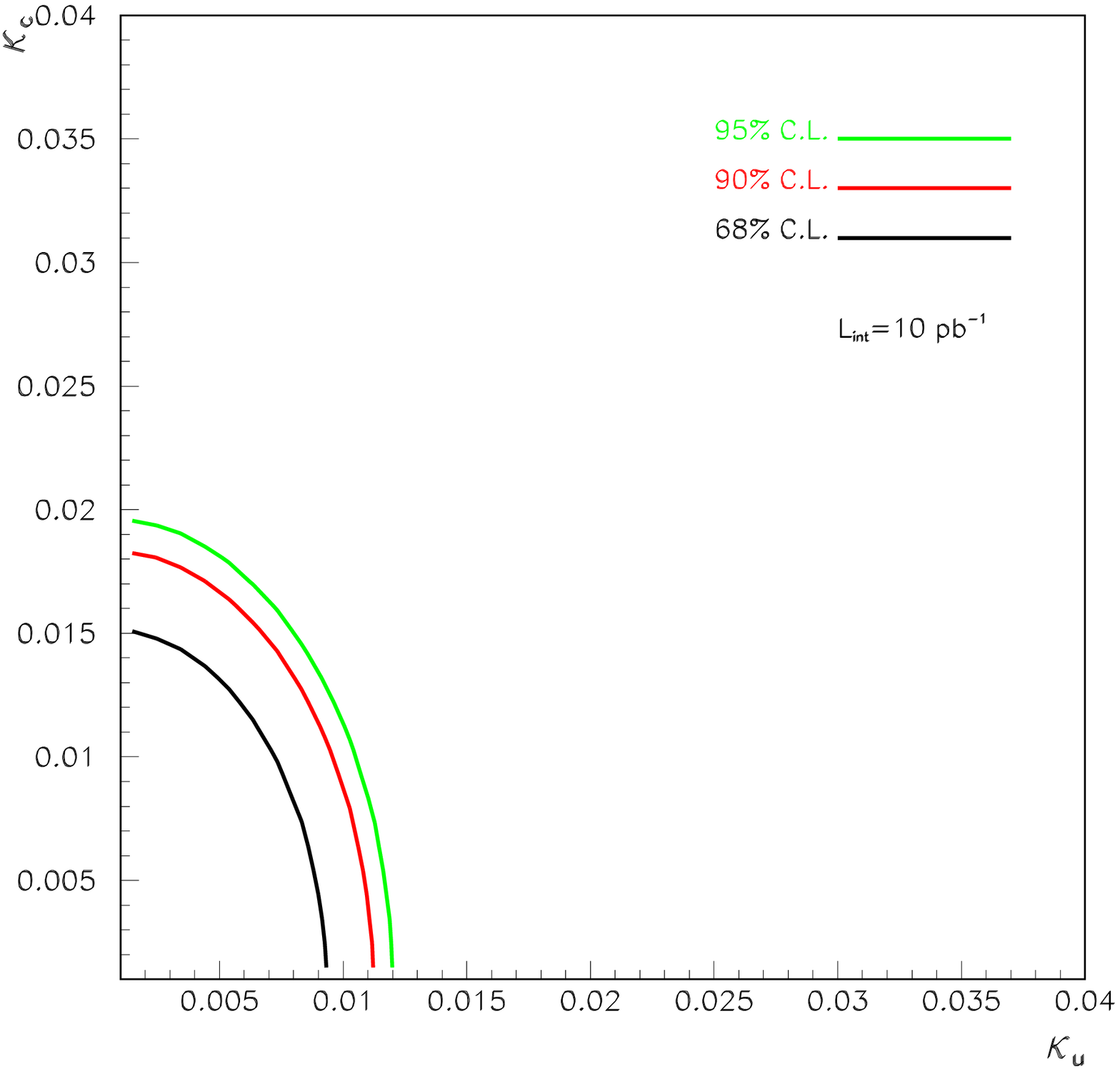}\caption{The same as Figure 7 but for the integrated luminosity $L_{int}=10$
fb$^{-1}$.\label{fig:8}}

\end{figure}

Up to now, we assume $\kappa_{u}=\kappa_{c}=\kappa$. However, it
is a matter of interest to analyze the $\kappa_{u}\neq\kappa_{c}$
case. Being different from HERA where anomalous single top production
is dominated by valence $u$-quarks, at LHeC energy region $c$-quark
contribution becomes comparable with the $u$-quark contribution.
Therefore, the sensitivity to $\kappa_{c}$ will be enhanced at LHeC
comparing to HERA. In Figs. \ref{fig:5}-\ref{fig:8} the contour
plots for the anomalous couplings in $\kappa_{u}-\kappa_{c}$ plane
are presented. For this purpose, we perform a $\chi^{2}$ analysis
by using \begin{eqnarray}
\chi^{2} & = & \sum_{i=1}^{N}\left({\scriptstyle {\textstyle \frac{\mbox{\ensuremath{\sigma}}_{S+B}^{i}-\mbox{\ensuremath{\sigma}}_{B}^{i}}{\Delta\sigma_{B}^{i}}}}\right)^{2}\label{eq:4}\end{eqnarray}
where $\mbox{\ensuremath{\sigma}}_{B}^{i}$ is the cross-section for
the SM background in the $i^{th}$ bin. It includes both $b$-jet
and light-jet contributions with the corresponding efficiency factors.
In the $\sigma_{S+B}$ calculations, we take into account $\kappa_{u}$
different from $\kappa_{c}$ case as well as signal-background interference.
One can see from Figs. \ref{fig:5}-\ref{fig:8} that sensitivity
is enhanced by a factor of 1.5 when the luminosity changes from 2
fb$^{-1}$ to 10 fb$^{-1}$. Concerning the energy upgrade, increasing
electron energy from 70 GeV to 140 GeV results in 20$\%$ improvement
for $\kappa_{c}$. Increasing electron energy further (energy frontier
$ep$ collider) does not give essential improvement in sensitivity
to anomalous couplings \cite{17}.

Finally, we compare our results with the LHC potential. The value
of $\kappa/\Lambda=0.01$ TeV$^{-1}$ corresponds to $BR(t\rightarrow\gamma u)\approx2\times10^{-6}$
which is two orders smaller than the LHC reach with 100 fb$^{-1}.$
It is obvious that even upgraded LHC will not be competitive with
LHeC based $\gamma p$ collider in the search for anomalous $t\gamma q$
interactions. Different extensions of the SM (supersymmetry, little
Higgs, extra dimensions, technicolor etc.) predict branching ratio
$BR(t\rightarrow\gamma q)$=$O$(10$^{-5}$), hence the LHeC will
provide opportunity to probe these models.
\begin{acknowledgments}
We are grateful to E. Perez and G. Ünel for useful discussions. \.{I}.T.Ç
would like to thank CERN Theory Division for the support and hospitality.
S.S. acknowledges the support from Turkish Atomic Energy Authority.\end{acknowledgments}

\end{document}